\documentclass[12pt]{article}
\usepackage[sort&compress,square,comma,numbers]{natbib}
\usepackage{amsmath,amssymb,graphicx,slashed,bm}
\usepackage{feynmp}
\usepackage{subfigure}
\usepackage{braket}
\usepackage{cancel}

\usepackage[
	colorlinks=true,
	citecolor=black,
	linkcolor=black,
	urlcolor=blue,
	hypertexnames=false]{hyperref}

\makeatletter
\def\endfmffile{%
	\fmfcmd{\p@rcent\space the end.^^J%
		end.^^J%
		endinput;}%
	\if@fmfio
	\immediate\closeout\@outfmf
	\fi
	\ifnum\pdfshellescape>\z@
	\immediate\write18{mpost \thefmffile}%
	\fi}
\makeatother
\newcommand{\arXiv}[2]{\href{http://arxiv.org/pdf/#1}{{\tt #2/#1}}}
\newcommand{\arXivold}[1]{\href{http://arxiv.org/pdf/#1}{{\tt #1}}}

\newcommand{\beq}{\begin{eqnarray}}% can be used as {equation} or  {eqnarray}
\newcommand{\eeq}{\end{eqnarray}}

\begin{document}
\begin{center} %TITLE HERE
{\huge \bf Geometrizing the Anomaly} 
\end{center}

\begin{center} % AUTHORS HERE

{\bf Joshua Newey}$^1$, {\bf John Terning}$^1$ and {\bf Christopher B. Verhaaren}$^2$ \\
\end{center}
\vskip 8pt
\begin{center} 
$^1${\it Center for Quantum Mathematics and Physics (QMAP)\\Department of Physics, University of California, Davis, CA 95616}\\
$^2${\it Department of Physics and Astronomy, Brigham Young University, \\ Provo, UT 84602, USA}
\end{center}

\vspace*{0.1cm}
\begin{center} 
{\tt 
 \href{mailto:jnewey@ucdavis.edu}{jnewey@ucdavis.edu}\,
 \href{mailto:jterning@gmail.com}{jterning@gmail.com}\,
 \href{mailto:verhaaren@physics.byu.edu}{verhaaren@physics.byu.edu}}

\end{center}

\centerline{\large\bf Abstract}
\begin{quote}
Recently a manifestly gauge invariant formalism for calculating amplitudes in quantum electrodynamics was outlined in which the field strength, rather than the gauge potential, is used as the propagating field. To demonstrate the utility of this formalism we calculate the axial and gauge anomalies explicitly in theories with both electrically and magnetically charged particles. Usually the gauge anomaly is identified as an amplitude that (in certain theories) fails to be gauge invariant, so it seems particularly enlightening to understand it in a manifestly gauge invariant formalism. We find that the three photon amplitude is still anomalous in these same theories because it depends explicitly upon the choice of the Stokes surface needed to couple the field strength to sources, so the gauge anomaly arises from geometric considerations.
\end{quote}

\newpage
\vspace{0.5cm}

\noindent  {\it ``which...doth neatly declare how nature geometrizeth, and observeth order in all things."}\\

\hspace*{10cm} ---Sir Thomas Browne~\cite{Browne:1658}
\begin{center} ------------------ \end{center}
\vspace{0.4cm}

%%%%%%%%%%%%%%%%%%%%%%%%%%%%%%%%%%%%%%%%%%%%%%%%
\section{Introduction\label{s.Intro}}

In classical field theories like electrodynamics and general relativity one often works with potentials that are mathematically convenient but a step removed from physical observables. Calculations can be made in one of an infinite variety of gauges while the final quantities must be independent of this gauge choice. Typically one chooses the gauge that simplifies the calculation and then relies on gauge invariance to ensure the generality of the result.

While potentials and gauges are simply a convenience in classical field theory, they have taken on a more essential role in quantum theories. Often, the work of Aharonov and Bohm \cite{Aharonov:1959fk} is given as evidence that the electromagnetic potential is essential to providing a local coupling to a charged particle. However recent work on the spinor helicity formalism~\cite{Elvang:2013cua} has emphasized the advantages of focussing on the helicity eigenstates of on-shell particles. In QED this means relying on the two physical degrees of freedom of the massless photon rather than the four gauge dependent degrees of freedom of the gauge potential. The price is that an arbitrary reference vector appears at intermediate steps of calculations. This most recent work builds upon a long history of research directed to working only with the field strength in QED~\cite{Dirac:1955uv,Mandelstam:1962mi,DeWitt:1962mg,Mandelstam:1968hz,Steinmann:1983ar} or to construct a gauge-independent potential formalism~\cite{Chen:2008ag,Lorce:2012rr}.

In a recent paper \cite{Newey:2024gug} we showed how the calculation of the perturbative running of the electric and magnetic couplings could be recast in terms of the gauge invariant field strength. The field strength does not couple directly to currents, but rather to a Stokes surface bounded by the worldline of the current. In this geometric picture one exchanges the ambiguity of gauge choice for the ambiguity of surface choice: the requirement of gauge invariance being exchanged for the requirement of surface independence. 
This proved crucial in distinguishing terms arising in the perturbation series which were related to overall topological phases which have often confused the extraction of the correct dynamics in such mutually non-local theories \cite{Terning:2018udc}. 
Around the same time, A related approach \cite{Hull:2024} was used to understand higher form symmetries of theories with monopoles.

That there exists an explicitly gauge invariant method for calculating in $U(1)$ gauge theories may seem too good to be true. And indeed, simplicity in regard to some aspects are traded for complications in others. Most notably, the field strength cannot couple directly to the conserved current, but instead couples to a surface bounded by the current worldline, essentially the Stokes surface of the eponymous theorem. But which surface should one use? As in classical electrodynamics, the physical results must be independent of the surface that is chosen. Thus, in using the field strength we have traded the requirement of gauge invariance for the geometrical requirement of surface independence.
The symmetry of the theory under changes in surface can be related to a one form gauge symmetry as described in \cite{Hull:2024}.
Here we see how this manifestly gauge invariant method can be applied to a notoriously gauge-non-invariant amplitude: the gauge anomaly.

In the following section we review the method for calculating in this field strength formalism, highlighting the calculation of amplitudes with external photons, which did not arise in our previous work. We also review the axial anomaly and show how the standard potential formalism and the field strength formalism are related in this case.  Then, in Sec.~\ref{Anomalies_Geometry}, we show how the axial anomaly and the three photon amplitude (gauge anomaly) are constrained by surface independence in the field strength formalism. We find that while the result for the three photon amplitude is explicitly gauge invariant, it depends on the choice of Stokes surface. Therefore, the requirement that all physical results are independent of surface choice leads to the usual anomaly constraint on $U(1)$ gauge theories. In Sec.~\ref{Cross_Checks} we calculate the axial anomaly for a particle with both electric and magnetic charge (a dyon) including the effects of a nonzero theta term. Our method quickly produces the gauge invariant result for all charge types which we compare the results to calculations based on duality. We also comment on how these results compare with calculations that use the Zwanziger formalism.  We conclude and offer an outlook on future work in Sec.~\ref{s.con}.

%%%%%%%%%%%%%%%%%%%%%%%%%%
\section{Review\label{s.review}}
In a classic work, Weinberg \cite{Weinberg:1965rz} showed the tight connection between massless spin-1 fields and Maxwell's equations. In so doing he states that the field strength cannot be used as the propagating field. The justification for this claim is that the components of the field strength's polarization tensor $\epsilon^h_{\mu\nu}(p)$  have the form
\beq
\epsilon^h_{\mu\nu}(p)= -ip_\mu\epsilon^h_{\nu}(p)+ ip_\nu\epsilon^h_{\mu}(p)~.
\label{eq.poltensor}
\eeq
where $\epsilon^h_{\mu}$ is the usual potential polarization vector with helicity $h$. This form does not give rise to long range forces, or even the Coulomb law, when locally coupled to a current. In our previous work~\cite{Newey:2024gug}, however, we do exactly obtain the correct long range interactions when using field strength propagation. This occurs because the couplings to currents are not local, but involve $1/(n\cdot p)$ interactions, where $n^\mu$ is the vector the describes the Stokes surface. 

Suppose we are interested in using the field strength as the propagating field. We might imagine the mode expansion of the source-free field strength as
\beq
F_{\mu\nu}(x)=\int \frac{d^3p}{(2\pi)^3}\frac{1}{\sqrt{2\omega_p}}\left(a_p^h\epsilon_{\mu\nu}^h e^{-ip\cdot x}+a^{\dag\,h}_p\overline{\epsilon}^h_{\mu\nu} e^{i p\cdot x} \right)~,
\eeq
where 
%$\epsilon_{\mu\nu}^h(p)$ is the $h$th component of the basis of polarization tensors and 
$\omega_p=|\vec{p}|$ and $\overline{\epsilon}^h_{\mu\nu} $ is the complex conjugate of $\epsilon^h_{\mu\nu} $. Because these polarization tensors arise as the coefficients of plane-wave solutions to the source-free Maxwell equations we have
\beq
p^\mu\epsilon_{\mu\nu}^h(p)=0~, \ \ \ \text{and} \ \ \ p_\mu\varepsilon^{\mu\nu\alpha\beta}\epsilon_{\alpha\beta}^h(p)=0~.
\label{eq:Fpolar}
\eeq
The second constraint comes by way of the dual field strength
\beq
{}^\ast\!F_{\mu\nu}(x)=\frac12\varepsilon^{\mu\nu\alpha\beta}F_{\alpha\beta}(x)=\int \frac{d^3p}{(2\pi)^3}\frac{1}{\sqrt{2\omega_p}}\left(a_p^h{}^\ast\!\epsilon_{\mu\nu}^h e^{-ip\cdot x}+a^{\dag\,h}_p{}^\ast\overline{\epsilon}^h_{\mu\nu} e^{i p\cdot x} \right)~,
\eeq
where
\beq
{}^\ast\!\epsilon^{h\mu\nu }=\frac12\varepsilon^{\mu\nu\alpha\beta}\epsilon_{\alpha\beta}^h~.\label{e.ep2epstar}
\eeq
This relation can be equivalently expressed as
\beq	
\epsilon^{ h \mu\nu}=-\frac12\varepsilon^{\mu\nu\alpha\beta}{}^\ast\!\epsilon_{\alpha\beta}^h~.\label{e.epstar2ep}
\eeq

\begin{figure}
\centering
\begin{tabular}{cc}
\begin{tabular}{c}
\begin{fmffile}{treePotFeyn}
\begin{fmfgraph*}(100,75)
\fmfpen{1.0}
\fmfstraight
\fmfset{arrow_len}{3mm}
\fmfleft{p1,i1,p2} \fmfright{o1,p3,o2}
\fmf{photon,tension=0.9}{i1,v1}
\fmf{fermion,tension=0.8}{o1,v1,o2}
\fmfv{decor.shape=circle,decor.filled=full,decor.size=1.5thick,l=$e$}{v1} 
\fmfv{l=$A_\mu$}{i1} 
\fmfv{l=$J^\mu$}{o1} 
\fmfv{l=$J^\mu$}{o2} 
\end{fmfgraph*}
\end{fmffile}
\end{tabular}
\begin{tabular}{c}
$\phantom{AA}-ie\gamma^\mu$
\end{tabular}
\end{tabular}
\caption{\label{f.potFeynRule}Standard Feynman rule (for a Dirac fermion) in potential formalism.}
\end{figure}
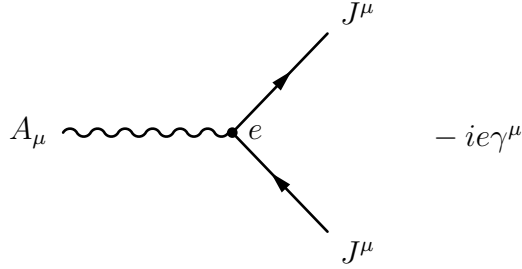

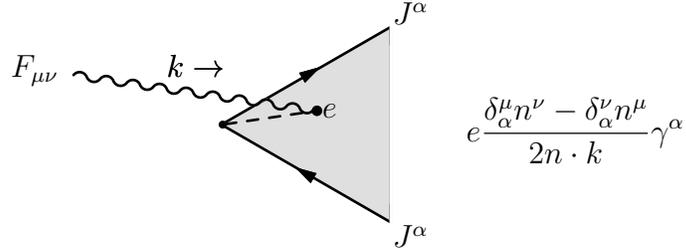
\begin{figure}
\centering
\begin{tabular}{cc}
\begin{tabular}{c}
\begin{fmffile}{treeFSFeyn}
\begin{fmfgraph*}(120,75)
\fmfpen{1.0}
\fmfstraight
\fmfset{arrow_len}{3mm}
\fmfleft{p4,p1,p2,i1,p5} \fmfright{o1,p3,o2}
\fmf{phantom,tension=0.2}{p2,v1}
\fmf{phantom,tension=0.922}{v1,p3}
\fmf{plain,tension=0.9,fore=(1,,1,,1),rubout=0.9}{o1,o2}
\fmffreeze
\fmf{photon,tension=0.5,rubout=0.9,label=$k\rightarrow$,label.side=left}{i1,v2}
\fmf{phantom,tension=0.85}{o2,v2}
\fmf{phantom,tension=0.8}{o1,v2}
\fmffreeze
\fmf{phantom,tension=0.53}{p2,v3}
\fmf{phantom,tension=0.47}{v3,p3}
\fmf{dashes,tension=0.0,rubout=0.9}{v3,v2}
\fmf{phantom_arrow,tension=0.0,rubout=0.9}{o1,v3,o2}
\fmfv{decor.shape=triangle,decor.size=85,decoration.angle=-30,background=(0.875,,0.875,,0.875),fore=(0.0,,0.0,,0.0),decor.fill=0}{v1} 
\fmfv{decor.shape=circle,decor.filled=full,decor.size=1.5thick,fore=(0.0,,0.0,,0.0),l=$e$,l.a=0,l.d=2}{v2} 
\fmfv{decor.shape=circle,decor.filled=full,decor.size=thick,fore=(0.0,,0.0,,0.0)}{v3} 
\fmfv{l=$F_{\mu\nu}$}{i1} 
\fmfv{l=$J^\alpha$,l.a=-45,l.d=1}{o1} 
\fmfv{l=$J^\alpha$,l.a=45,l.d=1}{o2} 
\end{fmfgraph*}
\end{fmffile}
\end{tabular}
\begin{tabular}{c}
$\phantom{A}\displaystyle e\frac{\delta^\mu_\alpha n^\nu-\delta^\nu_\alpha n^\mu}{2n\cdot k}\gamma^\alpha$
\end{tabular}
\end{tabular}
\caption{\label{f.fsFeynRule} Feynman rule (for a Dirac fermion) in field strength formalism. The shaded region indicates the Stokes surface bounded by the current. The field strength couples to the surface, the dashed line indicates the $1/n\cdot k$ propagation along the surface from the current. Note that the sign depends on whether the photon momentum is going into the current or out of the current.}
\end{figure}

What then does the Lagrangian look like? What are the Feynman rules that should be used? The electric source term in the Lagrangian, when using the gauge potential, is
\beq
-eA_\mu J^\mu~,
\eeq
and leads to the standard Feynman rule given in Fig.~\ref{f.potFeynRule}. In the field strength formalism \cite{Newey:2024gug} we find the Lagrangian coupling  term
\beq
eF_{\mu\nu}\frac{J^\mu n^\nu-J^\nu n^\mu}{2n\cdot\partial}~,\label{e.Fcoup}
\eeq
where the inverse derivative operator is understood to act on the currents, just as in Dirac's non-local formulation \cite{Dirac1948} of QED with mutually non-local charges. The appearance of non-local couplings may be somewhat unfamiliar, but recent work \cite{Cohen:2024fak} has shown that they can be quite useful when employed with care. This interaction leads to the momentum space Feynman rule given in Fig.~\ref{f.fsFeynRule}. We see that the field strength does not couple to the current, but rather to the surface (the gray shaded region) bounded by the current. The nonlocal part of the coupling is effectively a surface propagator (shown as a dashed line) from the point of the current's change in momentum to the point on the surface that the photon couples to. It is also significant that, unlike the potential interaction, the sign in the Feynman rule depends on whether the momentum in the photon is flowing into the current or out of the current. 

Using this Feynman rule we can often transform QED results in the potential formalism immediately into the field strength formalism. For instance, in the potential formalism we might have a fermionic amplitude (assumed for simplicity to have no internal photons) with two external, final-state photons of momenta $p^\mu$ and $k^\mu$ 
\beq
\mathcal{M}=\mathcal{M}^{\mu\nu}(p,k)\overline{\epsilon}_{\mu}(p)\overline{\epsilon}_{\nu}(k)~,
\eeq
where $\overline{\epsilon}_\mu(p)$ denotes the complex conjugate of the polarization vector. 
This amplitude can be rewritten in terms of field strength polarization tensors as
\beq
\mathcal{M}=\mathcal{M}^{\alpha\beta}(p,k)\left( \frac{1}{i}\right)^2\frac{\delta^\mu_\alpha n^\nu-\delta^\nu_\alpha n^\mu}{2n\cdot p}\overline{\epsilon}_{\mu\nu}(p)\frac{\delta^\sigma_\beta n^\rho-\delta^\rho_\beta n^\sigma}{2n\cdot k}\overline{\epsilon}_{\sigma\rho}(k)~.
\eeq
Here the factors of $i$ appear because the respective Feynman rules differ by a factor of $i$. In this case the signs are the same because the photons are outgoing from the fermion amplitude. The important point here is that $\mathcal{M}^{\mu\nu}(p,k)$ is the same in each case. This means that many calculated sub-amplitudes, like particle loops, that have been evaluated for potential formalism calculations need not be recomputed when moving to the field strength formalism.

%%%%%%%%%%%%%%
\subsection{The Axial Anomaly}
\label{Axial_Anomaly}
\begin{figure}[h]
\centering
\begin{fmffile}{chiAnom2pho}
\begin{fmfgraph*}(120,75)
\fmfpen{1.0}
\fmfstraight
\fmfset{arrow_len}{3mm}
\fmfleft{p1,i1,p2} \fmfright{o1,p3,o2}
\fmf{phantom,tension=0.15}{p1,v3}
\fmf{photon,tension=0.5}{v3,o1}
\fmf{phantom,tension=0.15}{p2,v2}
\fmf{photon,tension=0.5}{v2,o2}
\fmf{phantom,tension=0.5}{i1,v1}
\fmf{phantom,tension=0.2}{v1,p3}
\fmffreeze
\fmf{fermion,tension=0.1}{v1,v2,v3,v1}
\fmfv{decor.shape=circle,decor.size=0.5,l=$\bm{\otimes}$,l.a=0,l.d=0}{v1}
\fmfv{decor.shape=circle,decor.filled=full,decor.size=1.5thick,l=$eq$,l.a=180,l.d=6}{v2}
\fmfv{decor.shape=circle,decor.filled=full,decor.size=1.5thick,l=$eq$,l.a=180,l.d=6}{v3}
%\fmfv{l=$(p+k)^\alpha$}{i1} 
\fmfv{l=$k^\nu$,l.a=0,l.d=3}{o1} 
\fmfv{l=$p^\mu$,l.a=0,l.d=3}{o2} 
\end{fmfgraph*}
\end{fmffile}
\caption{\label{f.ChiAnom} A triangle diagram that contributes to the axial anomaly; the cross indicates the axial current insertion.  Note that the full amplitude  includes a second diagram in which the photons are exchanged.}
\end{figure}
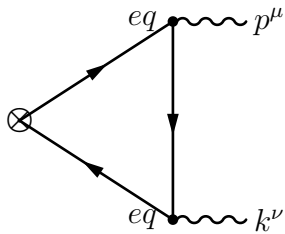

In this section we briefly remind the reader of the standard calculation of the axial anomaly in the potential formalism. The axial anomaly is obtained by taking the correlator of the axial current $J^\lambda_A$ and two gauge potentials. Thus we calculate triangle diagrams of the form
given in Fig.~\ref{f.ChiAnom}. 
The well known result is that the divergence of the axial current is given by  external momentum of the axial current $i(p+k)_\lambda$ contracted into the current vertex of the triangle diagram, labeled with a cross in Fig.~\ref{f.ChiAnom}. The resulting amplitude for particles with electric charge $q$ is
\begin{align}
\mathcal{M}=&\frac{e^2q^2}{2\pi^2}\varepsilon^{\alpha\mu\beta\nu}k_\alpha \overline{\epsilon}_{\mu}(k) p_\beta \overline{\epsilon}_{\nu}(p)\nonumber\\
=&\frac{-e^2q^2}{4\pi^2}\frac12\varepsilon^{\alpha\mu\beta\nu}\left[ik_\alpha \overline{\epsilon}_{\mu}(k)-ik_\mu \overline{\epsilon}_{\alpha}(k) \right]\left[ip_\beta \overline{\epsilon}_{\nu}(p)-ip_\nu \overline{\epsilon}_{\beta}(p) \right]~.
\end{align}
We can write this in terms of the field strength polarization tensors using \eqref{eq.poltensor}:
\beq
\mathcal{M}=\frac{-e^2q^2}{4\pi^2}\overline{\epsilon}_{\mu\nu}(k){}^\ast\overline{\epsilon}^{\mu\nu}(p)~.\label{e.chiralPot}
\eeq

Let us compare this with the result of applying the field strength formalism Feynman rule shown in Fig. \ref{f.fsFeynRule} which gives
\beq
\mathcal{M}=\mathcal{M}_\chi^{\mu\nu}(k,p)\frac{e^2q^2}{i^2}\frac{\delta_\mu^\gamma n^\delta-\delta_\mu^\delta n^\gamma}{2n\cdot k}\overline{\epsilon}_{\gamma\delta}(k)\frac{\delta_\nu^\sigma n^\rho-\delta_\nu^\rho n^\sigma}{2n\cdot p}\overline{\epsilon}_{\sigma\rho}(p)~,\label{e.Famp}
\eeq
because both photon lines are outgoing. From the potential based calculation above, the fermion loop part of the amplitude is simply
\beq
\mathcal{M}_\chi^{\mu\nu}(k,p)=\frac{1}{2\pi^2}\varepsilon^{\alpha\mu\beta\nu}k_\alpha p_\beta~.\label{e.triAmp}
\eeq
By inserting $\mathcal{M}_\chi^{\mu\nu}(k,p)$ into the field strength bases amplitude in Eq.~\eqref{e.Famp} and using the antisymmetric form of the polarization tensors we find
\beq
\mathcal{M}=-\frac{e^2q^2}{2\pi^2}\frac{\varepsilon^{\alpha\mu\beta\nu}k_\alpha p_\beta n^\delta n^\rho}{(n\cdot k)(n\cdot p)}\overline{\epsilon}_{\mu\delta}(k)\overline{\epsilon}_{\nu\rho}(p)~.
\eeq 
While this result does not look like the standard result it can be put into the usual form without much trouble. All one must do is write the polarization tensors in terms of their duals and use the usual identities for contracting Levi-Civita tensors.

In this case, we first write the $\overline{\epsilon}_{\mu\delta}(k)$ tensor and  the $\overline{\epsilon}_{\nu\rho}(p)$ tensor in terms of  their duals using \eqref{e.epstar2ep}. After contracting Levi-Civita tensors  we finally take the $k$ dependent tensor back from the dual. We find
\begin{align}
\mathcal{M}=&\frac{e^2q^2}{4\pi^2}\frac{ n^\rho n^\lambda}{(n\cdot k)(n\cdot p)}\varepsilon^{\alpha\mu\beta\nu}k_\alpha p_\beta\varepsilon_{\mu\rho\gamma\delta}{}^{\ast\,}\!\overline{\epsilon}^{\,\gamma\delta}(k)\overline{\epsilon}_{\nu\lambda}(p)\nonumber\\
=&-\frac{e^2q^2}{4\pi^2}\frac{ n^\rho n^\lambda}{(n\cdot k)(n\cdot p)}\varepsilon^{\mu\alpha\beta\nu}k_\alpha p_\beta\varepsilon_{\mu\rho\gamma\delta}{}^{\ast\,}\!\overline{\epsilon}^{\,\gamma\delta}(k)\overline{\epsilon}_{\nu\lambda}(p)\nonumber\\
=&\frac{e^2q^2}{2\pi^2}\frac{n^\lambda p_\gamma}{n\cdot p}{}^{\ast\,}\!\overline{\epsilon}^{\,\gamma\nu}(k)\overline{\epsilon}_{\nu\lambda}(p)=-\frac{e^2q^2}{8\pi^2}\frac{n^\lambda p_\gamma}{n\cdot p}\overline{\epsilon}_{\alpha\beta}(k)\varepsilon^{\gamma\nu\alpha\beta}{}^{\ast\,}\overline{\epsilon}^{\,\sigma\rho}(p)\varepsilon_{\nu\lambda\sigma\rho}\nonumber\\
=&-\frac{e^2q^2}{4\pi^2}\overline{\epsilon}_{\mu\nu}(k){}^{\ast\,}\!\overline{\epsilon}^{\,\mu\nu}(p)~.
\label{eq:FieldAxialAnom}
\end{align}
This exactly matches the standard result~\eqref{e.chiralPot}, as it must. We note that the process of going from dual polarization tensor and back has the effect of eliminating all $n^\mu$ dependence from the result. This illustrates the important point that amplitudes may appear to depend on Stokes surface reference vector, $n^\mu$, when in actuality they do not. 

The standard result~\eqref{e.chiralPot}, gives  the Fourier transform of the anomalous divergence of the axial current:
\beq
i(p+k)_\lambda \langle p,\,k| J^\lambda_A|0\rangle= -\frac{e^2q^2}{8\pi^2}\langle p,\,k|\frac12\varepsilon^{\alpha\mu\beta\nu}F_{\alpha\mu}F_{\beta\nu}|0\rangle~,
\eeq
where the additional factor of 2 on the right comes from connecting the gauge fields to the final state which is symmetric under the interchange of photons.

%%%%%%%%%%%%%%%%%%%%%%%%%%%%%%%%%%%%%%%%%%%%%%%%%
%%%%%%%%%%%%%%%%%%%%%%%%%%%%%%%%%%%%%%%%%%%%%%%%%
\section{Anomalies and Surface Geometry}
\label{Anomalies_Geometry}
In this section we clarify how gauge anomalies should be understood in the field strength formalism. Because the field strength is a gauge invariant object, no gauge redundancy appears at any point during the calculation. However, there is a redundancy in the choice of Stokes surface which the field strength couples to. Given the mathematical fact that all surfaces which span the same current worldline \cite{Strassler:1992zr,Schubert:1996jj} are equivalent, physical results must be independent of the surface choice. 

In the field strength formalism momentum space calculations are most easily performed by choosing  a fixed reference four vector, $n^\mu$, and the surface is spanned by a positive real number times $n^\mu$ attached to all points on the world-line. As shown below, the anomalies which arise in standard gauge theory as non-conserved currents show up in this formalism through a dependence on the choice of surface. 

This structure is reminiscent of, but distinct from, the way Dirac strings appear in theories with magnetic monopoles. In such theories a superficial dependence on the Dirac string may appear (due to topological terms which exponentiate to an overall phase of an amplitude \cite{Terning:2018udc,Newey:2024gug}). This dependence also includes $1/(n\cdot p)$ contributions, which can obscure the physical predictions of such theories~\cite{Terning:2020dzg}. However, these factors only appear in diagrams with internal photon lines connecting electric charges to magnetic charges. The relevant diagrams for anomalies do not contain such interactions.

The coupling in the field strength formalism defined in Eq.~\eqref{e.Fcoup} can be written as
\begin{align}
	\mathcal{L}_{\text{int}}
	&=
	eF^{\mu\nu}
	\frac{n_\mu J_\nu}{n\cdot \partial} ~.
\end{align}
The corresponding amplitude for a process with a single external photon 
\begin{align}
	\mathcal{M}
	=
	\epsilon^{\mu\nu}
	\frac{1}{n \cdot p}
	n_{\mu} M_\nu~.
\end{align}
Here $M_\nu$ is some amplitude which includes fields that make up the current $J_\nu$. This amplitude seems to depend on the vector $n^\mu$, but in Sec.~\ref{Axial_Anomaly} we saw that apparent $n^\mu$ dependence does not always mean actual dependence. What we need is a robust way of understanding the $n^\mu$ dependence of this amplitude, or in other words how the amplitude depends on our choice of coupling surface.

We begin by noting that any antisymmetric tensor $X_{\mu\nu}$ satisfies the identity~\cite{Newey:2024gug}
\begin{align}
	X_{\mu\nu}
	&=
	\frac{n^\alpha}{n\cdot p}
	(
	p_{\mu}
	X_{\alpha\nu}
	-
	p_{\nu}
	X_{\alpha\mu})
	-
	\frac{n^\alpha 
		p_\gamma}{2 n\cdot p}
	\epsilon_{\mu\nu\alpha\beta}
	\epsilon^{\gamma\beta\sigma\rho}
	X_{\sigma\rho} ~.
\end{align}
For the field strength polarizations $\epsilon_{\mu\nu}(p)$ the second term vanishes by the Bianchi identity. Therefore, any field strength polarization tensor can be written as 
\begin{align}
	\epsilon_{\mu\nu}(p)
	&=
	\frac{n^\alpha}{n\cdot p}
	\left[
	p_{\mu}
	\epsilon_{\alpha\nu}(p)
	-
	p_{\nu}
	\epsilon_{\alpha\mu}(p)\right]~.
	\label{eq:epsident}
\end{align}
A dependable way to determine if an amplitude depends on the choice of surface is to take a derivative of the amplitude with respect to the surface defining vector $n^\alpha$. Using the identity in Eq.~\eqref{eq:epsident} we find
\begin{align}
	\frac{\partial\;}
	{\partial n_\alpha}
	\mathcal{M}
	&=
	\epsilon^{\alpha\nu}
	\frac{i}{n \cdot p}
	M_\nu
	- 
	p^\alpha
	\epsilon^{\lambda\nu}
	\frac{i}{(n \cdot p)^2}
	n_{\lambda} M_\nu
	\notag\\
	&=
	\frac{n_\lambda}{(n\cdot p)^2}
	( p^\nu
	\epsilon^{\alpha\lambda}
	- p^\alpha \epsilon^{\nu\lambda})
	M_\nu
	- 
	p^\alpha
	\epsilon^{\lambda\nu}
	\frac{i}{(n \cdot p)^2}
	n_{\lambda} M_\nu
	\notag\\
	&=
	\frac{n_\lambda}{(n\cdot p)^2}
	\epsilon^{\alpha\lambda}
	p^\nu M_\nu~.
\end{align}
We see that in the gauge invariant formalism amplitudes are independent of $n^\mu$ so long as the Ward identity, $p^\nu M_\nu=0$ is satisfied. In the potential formalism, the Ward identity is closely tied to gauge transformations that might be applied to external photon fields. That is, the requirement that quantities like $M_\mu$ obey the Ward identity ensures that amplitudes like $\epsilon^\mu M_\mu$ are unaffected by gauge transformations of the potential ($\epsilon_\mu(p)\to\epsilon_\mu(p)+\kappa\,p_\mu$). The calculation above shows that in the field strength formalism the Ward identity ensures that amplitudes are unaffected by choosing a different Stokes surface.

%%%%%%%%%%%%%%%%%%%%%%%%
\subsection{The Axial Anomaly\label{ss.AxAnom}}
We can see how $n^\mu$ independence plays out in a nontrivial way by considering how the axial anomaly is manifest in the field strength formalism. In order to simplify the calculation we can choose an independent Stokes surface (associated with $n_1^\mu$ or $n_2^\mu$) for each photon. The axial current is a global current, so it does not require a Stokes surface. Thus, in the field strength formalism the standard axial current triangle diagram amplitude is
\begin{align}
	\mathcal{M}^\mu_A
	&=
	\langle J^{A\mu} J^\alpha J^\gamma \rangle_{k,p}
	n^\beta_1 n^\delta_2
	\frac{1}{n_1 \cdot k}
	\overline{\epsilon}_{\alpha\beta}(k)
	\frac{1}{n_2\cdot p}
	\overline{\epsilon}_{\gamma\delta}(p)~.\label{e.anomAmp}
\end{align}
If the axial current was conserved the Ward identity would produce
\begin{align}
	(k + p)_\mu
	\mathcal{M}^\mu_A
	&=
	(k + p)_\mu
	\langle J^{A\mu} J^\alpha J^\gamma \rangle_{k,p}
	n^\beta_1 n^\delta_2
	\frac{1}{n_1 \cdot k}
	\overline{\epsilon}_{\alpha\beta}(k)
	\frac{1}{n_2\cdot p}
	\overline{\epsilon}_{\gamma\delta}(p)
	= 0 ~,
\end{align}
which is clearly independent of both $n_1^\mu$ and $n_2^\mu$. The whole point of the anomaly, however, is that the Ward identity does not hold for the axial current: the current is not conserved. Though nonzero, we know the answer must still be independent of $n_1^\mu$ and $n_2^\mu$, but this is not obvious from Eq.~\eqref{e.anomAmp}.

As shown in many descriptions (for example~\cite{ryder1996quantum}), the three point triangle amplitude in momentum space is given by
\begin{align}
    \langle J^{A}_\mu 
    J_\kappa J_\lambda \rangle_{k,p}
    &= 2e^2 q^2 S_{\mu\kappa\lambda}(k,p,a)~,
\end{align}
where $S_{\mu\kappa\lambda}$ is 
\begin{align}
    S_{\mu\kappa\lambda}
     = -\frac{1}{(2\pi)^4}
     \int d^4q
     \frac{\text{Tr}[\gamma_k
     (\cancel{q}+\cancel{a}- \cancel{k})
     \gamma_\mu \gamma_5
     (\cancel{q} +\cancel{a}+ \cancel{p})
     \gamma_\lambda \cancel{q}
     ]}{(q+a-k)^2(q+a+p)^2q^2}~.
     \label{eq:naiveAmp}
\end{align}
In this definition the vector $a^\mu$ is an arbitrary shift of the loop momentum $q^\mu$. In terms of this arbitrary shift, the properties of $S_{\mu\kappa\lambda}(k,p)$ include
\begin{align}
    (k + p)^\mu S_{\mu\kappa\lambda}(k,p)
    &=  \frac{1}{8\pi^2}
    \epsilon_{\mu\kappa\lambda\nu} (k + p)^\mu a^\nu~,
    \\
     k^\kappa S_{\mu\kappa\lambda}(k,p)
     &= \frac{1}{8\pi^2}
     \epsilon_{\mu\kappa\lambda\nu} k^\kappa (p^\nu + a^\nu)~,
     \\
     p^\lambda S_{\mu\kappa\lambda}(k,p)
     &=
     - \frac{1}{8\pi^2}
     \epsilon_{\mu\kappa\lambda\nu}
     p^\lambda (k^\nu-a^\nu)~.
\end{align}
The first of these results is most clearly tied to the axial current not being conserved unless $a^\mu=k^\mu+p^\mu$. The other two show that the usual electric currents are not conserved for this choice of $a^\mu$. However, both the electric currents are conserved for $a^\mu=k^\mu-p^\mu$. For this choice of momentum shift the axial current is not conserved, but produces the famous anomaly given in Eq.~\eqref{e.chiralPot}.

Following the techniques of the previous section, we determine how the amplitude $\mathcal{M}_A^\mu$ depends on $n_1^\alpha$ and $n_2^\gamma$ by taking derivatives with respect to these vectors. Using Eq.~\eqref{eq:epsident} and the identities listed above for arbitrary $a^\mu$ we find
\begin{align}
	\frac{\partial\;}{\partial n_1^\alpha}
	\mathcal{M}^\mu_A
	&= 
	\frac{n_{1\lambda}}{(n_1 \cdot k)^2}
	k_\nu
	\langle J^{A\mu}
	J^\nu J^\gamma \rangle n^\delta_2
	\overline{\epsilon}^{\alpha\lambda}(k)
	\overline{\epsilon}_{\gamma\delta}(p)
	\notag\\
	&=
	\frac{n_{1\lambda}}{(n_1 \cdot k)^2}
	\left(
	\frac{1}{4\pi^2}
	k_\nu
	\epsilon^{\mu\nu\gamma\sigma}
	(p_\sigma + a_\sigma)
	\right)
	n^\delta_2
	\overline{\epsilon}^{\alpha\lambda}(p)
	\overline{\epsilon}_{\gamma\delta}
	 ~, \\
	\frac{\partial\;}{\partial n_2^\gamma}
	\mathcal{M}^\mu_A
	&=
	\frac{n_{2\lambda}}{(n_2 \cdot p)^2}
	p_\nu
	\langle J^{A\mu}
	J^\alpha J^\nu \rangle n^\beta_1
	\overline{\epsilon}_{\alpha\beta}(k)
	\overline{\epsilon}^{\gamma\lambda}(p)
	\notag\\
	&=
	\frac{n_{2\lambda}}{(n_2 \cdot p)^2}
	\left(
	\frac{1}{4\pi^2}
	p_\nu
	\epsilon^{\mu\alpha\nu\sigma}
	( a_\sigma-k_\sigma)
	\right)
	n^\beta_1
	\overline{\epsilon}_{\alpha\beta}(k)
	\overline{\epsilon}^{\gamma\lambda}(p)~.
\end{align}
These results show that there is a unique choice of $a^\mu$ which makes the all surface dependence vanish: $a^\mu = k^\mu - p^\mu$. This is also the choice that ensures the electric currents are conserved and produces the known result for the anomalous axial current. As promised, we find that the amplitude is independent of Stokes surface exactly when the Ward identity is satisfied for all currents associated with a gauge symmetry. In the next section this connection is made even stronger.

%%%%%%%%%%%%%%%%%%%%%%%%%%%%%%%%%%%%%%%%%%%%%%%%%
\subsection{The Gauge Anomaly }
We now turn to the gauge anomaly. It is interesting to understand how gauge anomalies can arise in a gauge invariant formalism since the usual focus is on the lack of gauge invariance. That is, there is no gauge invariant operator that the  triangle diagrams correspond to. This issue has also prevented the a complete understanding  of the gauge anomaly in the gauge invariant, spinor-helicity formalism. As one may expect from our earlier discussion, the manifestation of the anomaly in our, gauge invariant, field strength formalism is an inability to make the amplitude surface independent.

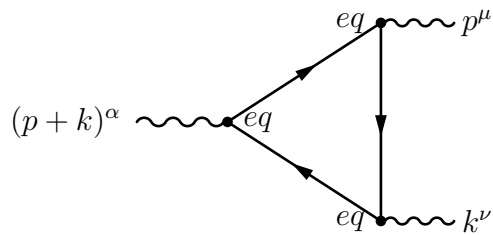
\begin{figure}[h]
\centering
\begin{fmffile}{Anom3pho}
\begin{fmfgraph*}(120,75)
\fmfpen{1.0}
\fmfstraight
\fmfset{arrow_len}{3mm}
\fmfleft{p1,i1,p2} \fmfright{o1,p3,o2}
\fmf{phantom,tension=0.15}{p1,v3}
\fmf{photon,tension=0.5}{v3,o1}
\fmf{phantom,tension=0.15}{p2,v2}
\fmf{photon,tension=0.5}{v2,o2}
\fmf{photon,tension=0.5}{i1,v1}
\fmf{phantom,tension=0.2}{v1,p3}
\fmffreeze
\fmf{fermion,tension=0.1}{v1,v2,v3,v1}
\fmfv{decor.shape=circle,decor.filled=full,decor.size=1.5thick,l=$eq$,l.a=0,l.d=6}{v1}
\fmfv{decor.shape=circle,decor.filled=full,decor.size=1.5thick,l=$eq$,l.a=180,l.d=6}{v2}
\fmfv{decor.shape=circle,decor.filled=full,decor.size=1.5thick,l=$eq$,l.a=180,l.d=6}{v3}
\fmfv{l=$(p+k)^\alpha$}{i1} 
\fmfv{l=$k^\nu$,l.a=0,l.d=3}{o1} 
\fmfv{l=$p^\mu$,l.a=0,l.d=3}{o2} 
\end{fmfgraph*}
\end{fmffile}
\caption{\label{f.gauge.Anom} A triangle diagram that contributes to gauge anomalies. Other diagrams are required to ensure photon symmetrization.}
\end{figure}

To demonstrate this we consider the simple example of standard QED with only (left-handed) Weyl fermions. In the standard formalism the potential couples to $J^\mu_L = \psi_L^\dagger \overline{\sigma}^\mu \psi_L = \overline{\psi
}\gamma^\mu P_L \psi$, and in the field strength formalism $F^{\mu\nu}$ couples to $\frac{-1}{2 (n\cdot\partial)} n^{[\mu} J_L^{\nu]}$. Following a standard derivation \cite{Schwartz:2014sze}, the three point Weyl current amplitude can be written as
\begin{align}
    \langle J^\mu_L J^\kappa_L J^\lambda_L \rangle_{k,p}
    = \frac{1}{2}
    \left(\langle J^\mu J^\kappa J^\lambda \rangle_{k,p}
    - \langle J^{A\mu} J^{\kappa} J^{\lambda} \rangle_{k,p}
    \right).
\end{align}
Where $\langle J^\mu J^\kappa J^\lambda \rangle_{k,p}$ is the standard vector three point amplitude which, in fact, vanishes by Furry's theorem, and $\langle J^{A\alpha} J^{\kappa} J^{\lambda} \rangle_{k,p}$ is the axial current amplitude we calculated in Sec.~\ref{ss.AxAnom}. The three point Weyl current amplitude in the field strength formalism can be written as
\begin{align}
	i\mathcal{M}_L
	&=
	-
	\frac{\epsilon_{\mu\nu}(p+k)
		\overline{\epsilon}_{\alpha\beta}(k)
		\overline{\epsilon}_{\gamma\delta}(p)
		n^\mu_1
		n^\alpha_2
		n^\gamma_3
		\langle J_L^\nu
		J_L^\beta J_L^\gamma \rangle}{\left[n_1\cdot(k+p)\right]
	(n_2\cdot k)
	(n_3\cdot p)~,
	}\label{e.gaugeAnomAmp}
\end{align}
where we have allowed for a different choice of reference vector $n_i^\mu$ for each external photon. As was the case with the axial anomaly, this is not strictly necessary, but allows us to consider the most general case.

Using the identity in \eqref{eq:epsident}  we find another manifestation of the surface independence  when the Ward identity is satisfied. Consider two vectors $n_\mu$ and $n'_\mu$. We can effectively exchange them by using Eq.~\eqref{eq:epsident}
\begin{align}
	\frac{1}{n\cdot p}
	\epsilon_{\mu\nu}
	n^\mu \langle J^\nu \mathcal{O}\rangle
	&= \frac{1}{(n \cdot p)(n' \cdot p)}
	(n')^\lambda
	(
	p_\nu \epsilon_{\mu\lambda}
	- p_\mu \epsilon_{\nu\lambda} 
	)
	n^\mu \langle J^\nu \mathcal{O}\rangle
	\notag\\
	&= 
	\frac{1}{n'\cdot p}
	\epsilon_{\lambda\nu}
	(n')^\lambda
	\langle J^\nu \mathcal{O}\rangle
	+ 
	\frac{1}{(n \cdot p)(n' \cdot p)}
	n^\mu
	(n')^\lambda
	\epsilon_{\mu\lambda}
	p_\nu 
	\langle J^\nu \mathcal{O}\rangle~\nonumber\\
	&=\frac{1}{n'\cdot p}\epsilon_{\lambda\nu}(n')^\lambda\langle J^\nu \mathcal{O}\rangle~,
\end{align}
where in the last step we have assumed the Ward identity causes the second term to vanish. 

Following the methods used above, we can determine how this amplitude in Eq.~\eqref{e.gaugeAnomAmp} depends on the $n_i^\mu$ by taking derivatives.  This immediately gives us three equations
\begin{align}
	\frac{\partial\;}{\partial n_1^\sigma}
	\mathcal{M}
	&=
	\frac{n^\lambda_1
	\epsilon_{\sigma\lambda}(p+k)
	\overline{\epsilon}_{\alpha\beta}(k)
	\overline{\epsilon}_{\gamma\delta}(p)
	n^\alpha_2
	n^\gamma_3
	}{\left[n_1\cdot(k+p)\right]^2
		(n_2\cdot k)
		(n_3\cdot p)
	}
	\left[
	(k+p)_\nu
	\langle J_L^\nu
	J_L^\beta J_L^\gamma \rangle
	\right]
	\notag\\
	&=
	\frac{n^\lambda_1
	\epsilon_{\sigma\lambda}(p+k)
	\overline{\epsilon}_{\alpha\beta}(k)
	\overline{\epsilon}_{\gamma\delta}(p)
	n^\alpha_2
	n^\gamma_3
	}{\left[n_1\cdot(k+p)\right]^2
		(n_2\cdot k)
		(n_3\cdot p)
	}
	\left[
	\frac{1}{8\pi^2}
	(k+ p)_\nu \epsilon^{\nu\beta\gamma\zeta}
	a_\zeta
	\right]~,\label{eq:Gaugendepend1}
	\\
	\frac{\partial\;}{\partial n_2^\sigma}
	\mathcal{M}
	&=
	\frac{n^\lambda_2
	\epsilon_{\mu\nu}(p+k)
	\overline{\epsilon}_{\sigma\lambda}(k)
	\overline{\epsilon}_{\gamma\delta}(p)
	n_1^\mu
	n^\gamma_3
	}{\left[n_1\cdot(k+p)\right]
		(n_2\cdot k)^2
		(n_3\cdot p)
	}
	\left[k_\beta
	\langle J_L^\nu
	J_L^\beta J_L^\gamma \rangle
	\right]
	\notag\\
	&=
	\frac{n^\lambda_2
	\epsilon_{\mu\nu}(p+k)
	\overline{\epsilon}_{\sigma\lambda}(k)
	\overline{\epsilon}_{\gamma\delta}(p)
	n_1^\mu
	n^\gamma_3
	}{\left[n_1\cdot(k+p)\right]
		(n_2\cdot k)^2
		(n_3\cdot p)
	}
	\left[
	\frac{1}{8\pi^2}
	k_\beta
	\epsilon^{\nu\beta\gamma\zeta}
	(p_\zeta  + a_\zeta)
	\right]~,\label{eq:Gaugendepend2}
	\\
	\frac{\partial\;}{\partial n_3^\sigma}
	\mathcal{M}
	&=
	\frac{n_3^\lambda
	\epsilon_{\mu\nu}
	(k+p)
	\overline{\epsilon}_{\alpha\beta}(k)
	\overline{\epsilon}_{\sigma\lambda}(p)
	n_1^\mu n_2^\alpha
	}{
	\left[n_1\cdot(k+p)\right]
	(n_2 \cdot k)
	(n_3 \cdot p)^2
	}
	\left[
	p_\gamma
	\langle J_L^\nu
	J_L^\beta J_L^\gamma \rangle
	\right]
	\notag\\
	&=
	\frac{n_3^\lambda
	\epsilon_{\mu\nu}
	(k+p)
	\overline{\epsilon}_{\alpha\beta}(k)
	\overline{\epsilon}_{\sigma\lambda}(p)
	n_1^\mu n_2^\alpha
	}{
		\left[n_1\cdot(k+p)\right]
		(n_2 \cdot k)
		(n_3 \cdot p)^2
	}
	\left[
	-\frac{1}{8\pi^2}
	p_\gamma
	\epsilon^{\nu\beta\gamma\zeta}
	(-k_\zeta
	+
	a_\zeta
	)
	\right]~.
	\label{eq:Gaugendepend3}
\end{align}
These results show that in order to have surface independence we must satisfy all three Ward identities. But, this is impossible. If we imagine choosing $a_\mu = \alpha k_\mu + \beta p_\mu$ then the derivatives in \eqref{eq:Gaugendepend1}\textendash\eqref{eq:Gaugendepend3} cannot all vanish for all polarizations and momenta. If they could then the momentum shift $a^\mu$ would be found bysolving the following system of simultaneous equations
\begin{align}
	\begin{cases}
		\alpha - \beta = 0 \\
		\beta + 1 = 0 \\
		-\alpha + 1 = 0 
	\end{cases}~.
\end{align}
These equations clearly have no solution.  Thus, unless the sum of cubed charges vanishes the gauge theory has amplitudes that depend on the choice of Stokes surface, which is anomalous. This result again emphasizes that while the potential formalism connects the Ward identity to gauge invariance the gauge invariant field strength formalism connects the Ward identity to Stokes surface independence.

%%%%%%%%%%%%%%%%%%%%%%%%%%%%%%%
\section{Application of the Field Strength Formalism}
\label{Cross_Checks}
An interesting application and cross check of the field strength formalism is to extend the familiar results discussed above to particles of all types of charge. That is, we calculate the axial anomaly for the cases of electric, magnetic, and dyonic\footnote{Including $\theta$ dependence via the Witten effect.} fermions. These types of calculations are 
generally very awkward in the standard potential formalism \cite{Ringwald}. In contrast, we can directly apply the same technique that was used in Sec.~\ref{Axial_Anomaly} to quickly determine the anomaly due to particles with other charges.

We begin with a fermion of magnetic charge $g$ running in the loop. The same diagram given in Fig.~\ref{f.ChiAnom} applies, with the simple substitution of $eq\to bg$, where $b$ represents the running coupling strength of the magnetic charge, while $q$ and $g$ satisfy the Dirac charge quantization condition \cite{Dirac:1931kp}. The diagram is interpreted in a slightly different way \cite{Newey:2024gug}, however. In this case the magnetic current $K^\mu$ couples directly to ${}^\ast F_{\mu\nu}$
\beq
b{}^\ast\!F_{\mu\nu}\frac{K^\mu n^\nu-K^\nu n^\mu}{2n\cdot\partial}~,
\eeq
 so the external photon lines in the diagram correspond to the dual of the polarization tensor. This leads to
\begin{align}
\mathcal{M}_b=&\mathcal{M}_\chi^{\mu\nu}(k,p)\frac{b^2g^2}{i^2}\frac{\delta_\mu^\gamma n^\delta-\delta_\mu^\delta n^\gamma}{2n\cdot k}{}^\ast\overline{\epsilon}_{\gamma\delta}(k)\frac{\delta_\nu^\sigma n^\rho-\delta_\nu^\rho n^\sigma}{2n\cdot p}{}^\ast\overline{\epsilon}_{\sigma\rho}(p)\nonumber\\
=&-\frac{b^2g^2}{2\pi^2}\frac{\varepsilon^{\alpha\mu\beta\nu}k_\alpha p_\beta n^\delta n^\rho}{(n\cdot k)(n\cdot p)}{}^\ast\overline{\epsilon}_{\gamma\delta}(k){}^\ast\overline{\epsilon}_{\sigma\rho}(p)~,
\end{align}
where $\mathcal{M}_\chi^{\mu\nu}(k,p)$ is result of the evaluating the fermion loop given in Eq.~\eqref{e.triAmp}. As in the electric case, we use Eqs.~\eqref{e.ep2epstar} and~\eqref{e.epstar2ep} to exchange the dual polarization tensors for the standard tensors and then trade one back to the dual. Using the standard Levi-Civita contraction identities we are able to eliminate $n^\mu$ and find
\begin{align}
\mathcal{M}_b=&\frac{b^2g^2}{4\pi^2}\frac{ n^\rho n^\lambda}{(n\cdot k)(n\cdot p)}\varepsilon^{\mu\alpha\beta\nu}k_\alpha p_\beta\varepsilon_{\mu\rho\gamma\delta}\overline{\epsilon}^{\,\gamma\delta}(k){}^{\ast\,}\!\overline{\epsilon}_{\kappa\lambda}(p)\nonumber\\
=&-\frac{b^2g^2}{2\pi^2}\frac{n^\lambda p_\gamma}{n\cdot p}\overline{\epsilon}^{\,\gamma\nu}(k){}^{\ast\,}\!\overline{\epsilon}_{\nu\lambda}(p)=\frac{b^2g^2}{8\pi^2}\frac{n^\lambda p_\gamma}{n\cdot p}{}^{\ast\,}\!\overline{\epsilon}_{\alpha\beta}(k)\varepsilon^{\gamma\nu\alpha\beta}\overline{\epsilon}^{\,\sigma\rho}(p)\varepsilon_{\nu\lambda\sigma\rho}\nonumber\\
=&\frac{b^2g^2}{4\pi^2}{}^{\ast\,}\!\overline{\epsilon}_{\mu\nu}(k)\overline{\epsilon}^{\,\mu\nu}(p)~.
\end{align}

We see that a magnetic fermion produces a nearly identical result to the electric fermion case. Importantly, it differs by contributing to the anomaly with opposite sign. One can think of this operationally as due to the initial epsilon tensor being contracted with one of the dual field strengths, producing a minus sign. This relative sign between electric and magnetic contributions also appears in the $\beta$-function coefficient for the running of the $e$ and $b$, as shown in~\cite{Newey:2024gug}. The was first emphasized by Coleman \cite{Coleman:1982cx}, argued for perturbatively by Panagiotakopoulos \cite{Panagiotakopoulos:1982ne} and confirmed in Seiberg-Witten \cite{SeibergWitten} theories by Argyres and Douglas \cite{Argyres:1995jj} and by an independent method in \cite{Colwell:2015wna}.

The magnetic calculation above appears isolated from the electric case. One can imagine it as pertaining to a theory with only magnetic charges and a perturbative magnetic coupling. However, this is not required for the calculation to have a meaningful interpretation. Kinetic mixing between two $U(1)$ theories with both types of charges~\cite{Terning:2018lsv} can lead to effective theories with perturbative couplings to both electric and magnetic particles~\cite{Brummer:2009cs,GomezSanchez:2011orv,Hook:2017vyc,Terning:2019bhg,Graesser:2021vkr}, in which case both would contribute to the anomaly. In addition, the one-loop exact nature of the axial anomaly suggests that even perturbative calculations can have some meaning for larger coupling values, though it is likely that Abelian instantons play an important role~\cite{Csaki:2024ajo}. In Seiberg-Witten~\cite{SeibergWitten} it is explicitly seen that monopole anomaly diagrams are dual to instanton sums.

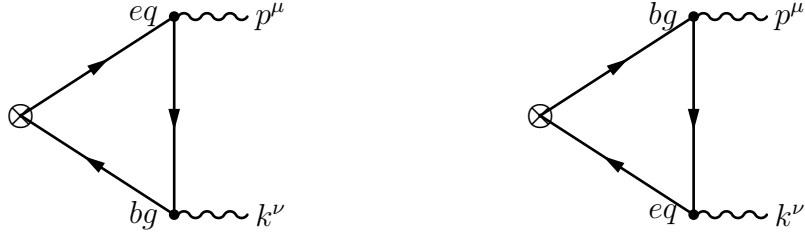
\begin{figure}
\centering
\begin{fmffile}{AnombUp}
\begin{fmfgraph*}(120,75)
\fmfpen{1.0}
\fmfstraight
\fmfset{arrow_len}{3mm}
\fmfleft{p1,i1,p2} \fmfright{o1,p3,o2}
\fmf{phantom,tension=0.15}{p1,v3}
\fmf{photon,tension=0.5}{v3,o1}
\fmf{phantom,tension=0.15}{p2,v2}
\fmf{photon,tension=0.5}{v2,o2}
\fmf{phantom,tension=0.5}{i1,v1}
\fmf{phantom,tension=0.2}{v1,p3}
\fmffreeze
\fmf{fermion,tension=0.1}{v1,v2,v3,v1}
\fmfv{decor.shape=circle,decor.size=0.5,l=$\bm{\otimes}$,l.a=0,l.d=0}{v1}
\fmfv{decor.shape=circle,decor.filled=full,decor.size=1.5thick,l=$eq$,l.a=180,l.d=6}{v2}
\fmfv{decor.shape=circle,decor.filled=full,decor.size=1.5thick,l=$bg$,l.a=180,l.d=6}{v3}
%\fmfv{l=$(p+k)^\alpha$}{i1} 
\fmfv{l=$k^\nu$,l.a=0,l.d=3}{o1} 
\fmfv{l=$p^\mu$,l.a=0,l.d=3}{o2} 
\end{fmfgraph*}
\end{fmffile}
\hspace{2cm}
\begin{fmffile}{AnombDown}
\begin{fmfgraph*}(120,75)
\fmfpen{1.0}
\fmfstraight
\fmfset{arrow_len}{3mm}
\fmfleft{p1,i1,p2} \fmfright{o1,p3,o2}
\fmf{phantom,tension=0.15}{p1,v3}
\fmf{photon,tension=0.5}{v3,o1}
\fmf{phantom,tension=0.15}{p2,v2}
\fmf{photon,tension=0.5}{v2,o2}
\fmf{phantom,tension=0.5}{i1,v1}
\fmf{phantom,tension=0.2}{v1,p3}
\fmffreeze
\fmf{fermion,tension=0.1}{v1,v2,v3,v1}
\fmfv{decor.shape=circle,decor.size=0.5,l=$\bm{\otimes}$,l.a=0,l.d=0}{v1}
\fmfv{decor.shape=circle,decor.filled=full,decor.size=1.5thick,l=$bg$,l.a=180,l.d=6}{v2}
\fmfv{decor.shape=circle,decor.filled=full,decor.size=1.5thick,l=$eq$,l.a=180,l.d=6}{v3}
%\fmfv{l=$(p+k)^\alpha$}{i1} 
\fmfv{l=$k^\nu$,l.a=0,l.d=3}{o1} 
\fmfv{l=$p^\mu$,l.a=0,l.d=3}{o2} 
\end{fmfgraph*}
\end{fmffile}
\caption{\label{f.dyonDiags} Contributions to $F_{\mu\nu}F^{\mu\nu}$ operator from dyonic fields. Each diagram is one of a pair, as in Fig.~\ref{f.ChiAnom}.}
\end{figure}

These field strength methods can also be immediately applied to dyonic fermions, particles with electric charge $q$ and magnetic charge $g$. In addition to the contributions obtained above, such fields produce diagrams in which one external photon couples with the electric charge to $F_{\mu\nu}$ with the other couples with magnetic charge to ${}^\ast\!F_{\mu\nu}$, see Fig.~\ref{f.dyonDiags}. Both choices for which external photon is either an electric coupling or a magnetic coupling produce the same result, leading to an overall factor of 2. 

We begin with
\begin{align}
\mathcal{M}_{eb}=&2\mathcal{M}_\chi^{\mu\nu}(k,p)\frac{eqbg}{i^2}\frac{\delta_\mu^\gamma n^\delta-\delta_\mu^\delta n^\gamma}{2n\cdot k}{}^\ast\overline{\epsilon}_{\gamma\delta}(k)\frac{\delta_\nu^\sigma n^\rho-\delta_\nu^\rho n^\sigma}{2n\cdot p}\overline{\epsilon}_{\sigma\rho}(p)\nonumber\\
=&-\frac{egbg}{\pi^2}\frac{\varepsilon^{\alpha\mu\beta\nu}k_\alpha p_\beta n^\delta n^\rho}{(n\cdot k)(n\cdot p)}{}^\ast\overline{\epsilon}_{\gamma\delta}(k)\overline{\epsilon}_{\sigma\rho}(p)~.
\end{align}
Then, as in the previous calculations, we then introduce factors of the epsilon tensor by exchanging polarization tensors with their duals and vice versa. 
\begin{align}
\mathcal{M}_{eb}=&-\frac{eqbg}{2\pi^2}\frac{ n^\rho n^\lambda}{(n\cdot k)(n\cdot p)}\varepsilon^{\mu\alpha\beta\nu}k_\alpha p_\beta\varepsilon_{\mu\rho\gamma\delta}\overline{\epsilon}^{\,\gamma\delta}(k)\overline{\epsilon}_{\kappa\lambda}(p)\nonumber\\
=&-\frac{eqbg}{\pi^2}\frac{n^\lambda p_\gamma}{n\cdot p}\overline{\epsilon}^{\,\gamma\nu}(k)\overline{\epsilon}_{\nu\lambda}(p)\nonumber\\
=&-\frac{eqbg}{4\pi^2}\frac{n^\lambda p_\gamma}{n\cdot p}{}^{\ast\,}\!\overline{\epsilon}_{\alpha\beta}(k)\varepsilon^{\gamma\nu\alpha\beta}{}^{\ast\,}\!\overline{\epsilon}^{\,\sigma\rho}(p)\varepsilon_{\nu\lambda\sigma\rho}=\frac{eqbg}{2\pi^2}{}^{\ast\,}\!\overline{\epsilon}_{\mu\nu}(k){}^{\ast\,}\!\overline{\epsilon}^{\,\mu\nu}(p)\nonumber\\
=&\frac{eqbg}{2\pi^2}\overline{\epsilon}_{\mu\nu}(k)\overline{\epsilon}^{\,\mu\nu}(p)~.
\end{align}
An important result of this calculation is that this is not a contribution to ${}^\ast\!F_{\mu\nu}F^{\mu\nu}$, but rather to $F_{\mu\nu}F^{\mu\nu}$. This is allowed (in fact required) because magnetic charges are odd under CP, see~\cite{Terning:2020dzg} for a more thorough discussion. Because the axial anomaly itself is CP odd the coefficient $eqbg$ must be paired with a CP even operator composed of two field strengths. In short, the total contribution to the axial anomaly from a fermion with electric or magnetic charge is
\beq
-\left[\frac{e^2q^2}{8\pi^2}-\frac{b^2g^2}{8\pi^2}\right]F_{\mu\nu}{}^\ast\!F^{\mu\nu}+\frac{eqbg}{4\pi^2}F_{\mu\nu}F^{\mu\nu}~.
\eeq

The relative ease with which these results were obtained encourages one to be even more general. Thus far we have not considered a nonzero $\theta$ term. As first shown by Witten~\cite{Witten:1979ey}, the inclusion of such a term requires that a particle with magnetic charge $g$ develops an electric charge $g\theta/(2\pi)$, which is distributed over a size determined by the Compton wavelength of the electrically charged fermion~\cite{Callan:1982ah}. At sufficiently low energy we can treat this as a local charge.  As shown in Figs.~\ref{f.OneTheta} and~\ref{f.TwoTheta} this produces additional sets of diagrams for a particle with electric charge $q$ and magnetic charge $g$. The new, $\theta$ dependent, electric charges contribute just like the $q$ charge, leading to a $\theta^2$ contribution to ${}^\ast\!F_{\mu\nu}F^{\mu\nu}$ and a $\theta g^2$ contribution to $F_{\mu\nu}F^{\mu\nu}$. In addition to these effects, we also have two mixed electric charge diagrams $\theta g q$ that contribute to ${}^\ast\!F_{\mu\nu}F^{\mu\nu}$.

\begin{figure}
\centering
\begin{fmffile}{AnomThUp}
\begin{fmfgraph*}(120,75)
\fmfpen{1.0}
\fmfstraight
\fmfset{arrow_len}{3mm}
\fmfleft{p1,i1,p2} \fmfright{o1,p3,o2}
\fmf{phantom,tension=0.15}{p1,v3}
\fmf{photon,tension=0.5}{v3,o1}
\fmf{phantom,tension=0.15}{p2,v2}
\fmf{photon,tension=0.5}{v2,o2}
\fmf{phantom,tension=0.5}{i1,v1}
\fmf{phantom,tension=0.2}{v1,p3}
\fmffreeze
\fmf{fermion,tension=0.1}{v1,v2,v3,v1}
\fmfv{decor.shape=circle,decor.size=0.5,l=$\bm{\otimes}$,l.a=0,l.d=0}{v1}
\fmfv{decor.shape=circle,decor.filled=full,decor.size=1.5thick,l=$eq$,l.a=180,l.d=6}{v2}
\fmfv{decor.shape=circle,decor.filled=full,decor.size=1.5thick,l=$eg\frac{\theta}{2\pi}$,l.a=180,l.d=12}{v3}
%\fmfv{l=$(p+k)^\alpha$}{i1} 
\fmfv{l=$k^\nu$,l.a=0,l.d=3}{o1} 
\fmfv{l=$p^\mu$,l.a=0,l.d=3}{o2} 
\end{fmfgraph*}
\end{fmffile}
\hspace{2cm}
\begin{fmffile}{AnomThDown}
\begin{fmfgraph*}(120,75)
\fmfpen{1.0}
\fmfstraight
\fmfset{arrow_len}{3mm}
\fmfleft{p1,i1,p2} \fmfright{o1,p3,o2}
\fmf{phantom,tension=0.15}{p1,v3}
\fmf{photon,tension=0.5}{v3,o1}
\fmf{phantom,tension=0.15}{p2,v2}
\fmf{photon,tension=0.5}{v2,o2}
\fmf{phantom,tension=0.5}{i1,v1}
\fmf{phantom,tension=0.2}{v1,p3}
\fmffreeze
\fmf{fermion,tension=0.1}{v1,v2,v3,v1}
\fmfv{decor.shape=circle,decor.size=0.5,l=$\bm{\otimes}$,l.a=0,l.d=0}{v1}
\fmfv{decor.shape=circle,decor.filled=full,decor.size=1.5thick,l=$eg\frac{\theta}{2\pi}$,l.a=180,l.d=12}{v2}
\fmfv{decor.shape=circle,decor.filled=full,decor.size=1.5thick,l=$eq$,l.a=180,l.d=6}{v3}
%\fmfv{l=$(p+k)^\alpha$}{i1} 
\fmfv{l=$k^\nu$,l.a=0,l.d=3}{o1} 
\fmfv{l=$p^\mu$,l.a=0,l.d=3}{o2} 
\end{fmfgraph*}
\end{fmffile}
\caption{\label{f.OneTheta} Contributions to $F_{\mu\nu}{}^\ast\!F^{\mu\nu}$ operator involving one instance of $\theta$. Each diagram is one of a pair, as in Fig.~\ref{f.ChiAnom}.}
\end{figure}

\begin{figure}
\centering
\begin{fmffile}{Anom2Th}
\begin{fmfgraph*}(120,75)
\fmfpen{1.0}
\fmfstraight
\fmfset{arrow_len}{3mm}
\fmfleft{p1,i1,p2} \fmfright{o1,p3,o2}
\fmf{phantom,tension=0.15}{p1,v3}
\fmf{photon,tension=0.5}{v3,o1}
\fmf{phantom,tension=0.15}{p2,v2}
\fmf{photon,tension=0.5}{v2,o2}
\fmf{phantom,tension=0.5}{i1,v1}
\fmf{phantom,tension=0.2}{v1,p3}
\fmffreeze
\fmf{fermion,tension=0.1}{v1,v2,v3,v1}
\fmfv{decor.shape=circle,decor.size=0.5,l=$\bm{\otimes}$,l.a=0,l.d=0}{v1}
\fmfv{decor.shape=circle,decor.filled=full,decor.size=1.5thick,l=$eg\frac{\theta}{2\pi}$,l.a=180,l.d=12}{v2}
\fmfv{decor.shape=circle,decor.filled=full,decor.size=1.5thick,l=$eg\frac{\theta}{2\pi}$,l.a=180,l.d=12}{v3}
%\fmfv{l=$(p+k)^\alpha$}{i1} 
\fmfv{l=$k^\nu$,l.a=0,l.d=3}{o1} 
\fmfv{l=$p^\mu$,l.a=0,l.d=3}{o2} 
\end{fmfgraph*}
\end{fmffile}
\caption{\label{f.TwoTheta} Contributions to $F_{\mu\nu}{}^\ast\!F^{\mu\nu}$ operator involving one instance of $\theta$. Each diagram is one of a pair, as in Fig.~\ref{f.ChiAnom}.}
\end{figure}

Putting all the terms together, we find that the total axial anomaly due to a dyon in the presence of a nonzero $\theta$ term is
\beq
-\left[\frac{e^2}{8\pi^2}\left(q+\frac{g\theta}{2\pi} \right)^2-\frac{b^2g^2}{8\pi^2}\right]F_{\mu\nu}{}^\ast\!F^{\mu\nu}+\frac{ebg}{4\pi^2}\left(q+\frac{g\theta}{2\pi} \right)F_{\mu\nu}F^{\mu\nu}~.
\label{eq.finalanomaly}
\eeq
This agrees with the results in~\cite{Csaki:2010rv} which employed the duality structure of QED to determine the form of the axial anomaly. In fact, our result is a useful check, in that we disagree with their published results by a factor of $-2$ on the $F_{\mu\nu}F^{\mu\nu}$ term. However, by applying their methods one finds that this is simply a typo. Thus, these two methods do in fact agree and provide independent checks on the results.

%%%%%%%%%%%%%%%%%%%
\subsection{Zwanziger Formalism}
An alternative approach to calculating the contribution of magnetic or dyonic particles to the axial anomaly is to use the Zwanziger formalism \cite{Zwanziger:1970hk}, which uses local couplings for both electric and magnetic charges but sacrifices manifest Lorentz invariance by involving a constant vector $n^\mu$. This formalism also requires two gauge potentials, $A^\mu$ and $B^\mu$, to describe the photon, and additional constraints to ensure that there are only two propagating degrees of freedom. This is the approach taken in refs.~\cite{Ringwald}. 

In this formalism the gauge invariant field strength is given by
\begin{align}
F_{\mu\nu}=&\frac{n^\alpha}{n^2}\left(n_\mu F^A_{\alpha\nu}-n_\nu F^A_{\alpha\mu}-\varepsilon_{\mu\nu\alpha\beta}n_\gamma F^{B\gamma\beta} \right),\\
{}^\ast\! F_{\mu\nu}=&\frac{n^\alpha}{n^2}\left(n_\mu F^B_{\alpha\nu}-n_\nu F^B_{\alpha\mu}+\varepsilon_{\mu\nu\alpha\beta}n_\gamma F^{A\gamma\beta} \right).
\end{align}
where
\beq
F^X_{\mu\nu}=\partial_\mu X_\nu-\partial_\nu X_\mu~.
\eeq
The potential $A^\mu$ couples locally to the electric $J^\mu$, while $B^\mu$ couples to the magnetic current $K^\mu$. We can also transform calculations in this formalism into the field strength formalism.
When the Zwanziger interactions are converted into the field strength coupling terms we find
\begin{align}
eF_{\mu\nu}\frac{J^\mu n^\nu-J^\nu n^\mu}{2n\cdot\partial}=&eF^A_{\mu\nu}\frac{J^\mu n^\nu-J^\nu n^\mu}{2n\cdot\partial}~,\\
 b{}^\ast\!F_{\mu\nu}\frac{K^\mu n^\nu-K^\nu n^\mu}{2n\cdot\partial}=&bF^B_{\mu\nu}\frac{K^\mu n^\nu-K^\nu n^\mu}{2n\cdot\partial}~.
\end{align}
We emphasize that in the field strength formalism the couplings can be expressed in terms of $F^A_{\mu\nu}$ and $F^B_{\mu\nu}$ or in terms of the physical field strength and its dual.

Therefore, we can perform the same calculations as above, but take electric charges as coupling to $F^A_{\mu\nu}$ and magnetic charges as coupling to $F^B_{\mu\nu}$. 
This allows us to determine the axial anomaly using the Zwanziger field strengths. Using the $F^A_{\mu\nu}$ coupling we obtain
\beq
-\frac{e^2q^2}{4\pi^2}\overline{\epsilon}^A_{\mu\nu}(k){}^\ast\overline{\epsilon}^{A\mu\nu}(p)~.\label{e.ZwanA}
\eeq
By a nearly identical calculation, but for a magnetic fermion, we find
\beq
-\frac{b^2g^2}{4\pi^2}\overline{\epsilon}^B_{\mu\nu}(k){}^\ast\overline{\epsilon}^{B\mu\nu}(p)~.\label{e.ZwanB}
\eeq
Notice that unlike the magnetic particle calculation made previously this has the same sign as the electric case, although the type of polarization tensor is different. 

For a dyonic field we have both of the results above as well as diagrams that include one external $F^A_{\mu\nu}$ and one $F^B_{\mu\nu}$. Because there are two way of associating the external photons with these two fields, this contribution includes an additional factor of two. The end result is
\beq
-\frac{ebqg}{2\pi^2}\overline{\epsilon}^A_{\mu\nu}(k){}^\ast\overline{\epsilon}^{B\mu\nu}(p)~,
\eeq
which again has the same sign as~\eqref{e.ZwanA} and~\eqref{e.ZwanB}. How should we interpret these results? One could, as was done in~\cite{Ringwald}, take them to correspond to the operators
\beq
-\frac{e^2q^2}{8\pi^2}F^A_{\mu\nu}{}^\ast\! F^{A\mu\nu}-\frac{b^2g^2}{8\pi^2}F^B_{\mu\nu}{}^\ast\! F^{B\mu\nu}-\frac{ebqg}{4\pi^2}F^A_{\mu\nu}{}^\ast\! F^{B\mu\nu}~,
\eeq 
which, on the face of it, does not appear to agree with \eqref{eq.finalanomaly}. This apparent conflict is resolved by making a careful examination, in the following subsection, of how the polarization tensors for the Zwanziger field strengths relate to the polarization tensors of the physical field strength. 

%%%%%%%%%%%%%%%%%%%%%%%%%
\subsection{Polarization Tensor Relations}
In the Zwanziger formalism the relationship between $F_{\mu\nu}$ both of $F^A_{\mu\nu}$ and $F^B_{\mu\nu}$ also applies to their polarization tensors:
\begin{align}
\epsilon_{\mu\nu}=&\frac{n^\alpha}{n^2}\left(n_\mu \epsilon^A_{\alpha\nu}-n_\nu \epsilon^A_{\alpha\mu}-\varepsilon_{\mu\nu\alpha\beta}n_\gamma \epsilon^{B\gamma\beta} \right),\\
{}^\ast\! \epsilon_{\mu\nu}=&\frac{n^\alpha}{n^2}\left(n_\mu \epsilon^B_{\alpha\nu}-n_\nu \epsilon^B_{\alpha\mu}+\varepsilon_{\mu\nu\alpha\beta}n_\gamma \epsilon^{A\gamma\beta} \right).
\end{align}
By using the definitions
\beq
\epsilon^{\mu\nu}=-\frac12\varepsilon^{\mu\nu\alpha\beta}{}^\ast\!\epsilon_{\alpha\beta}~,
\eeq
for both of the $A$ and $B$ polarization tensors we find that
\begin{align}
\epsilon_{\mu\nu}=&\frac{n^\alpha}{n^2}\left[n_\mu\left(\epsilon^A_{\alpha\nu}+{}^\ast\!\epsilon^B_{\alpha\nu}\right)-n_\nu\left(\epsilon^A_{\alpha\mu}+{}^\ast\!\epsilon^B_{\alpha\mu}\right)\right]-{}^\ast\!\epsilon^{B}_{\mu\nu}~,\label{e.FepsDef}\\
 {}^\ast\!\epsilon_{\mu\nu}=&\frac{n^\alpha}{n^2}\left[n_\mu\left(\epsilon^B_{\alpha\nu}-{}^\ast\!\epsilon^A_{\alpha\nu}\right)-n_\nu\left(\epsilon^B_{\alpha\mu}-{}^\ast\!\epsilon^A_{\alpha\mu}\right)\right]+{}^\ast\!\epsilon^{A}_{\mu\nu}~.\label{e.starFepsDef}
\end{align}

All of these polarization tensors satisfy
\beq
k^\mu\epsilon_{\mu\nu}(k)=k^\mu\epsilon^A_{\mu\nu}(k)=k^\mu\epsilon^B_{\mu\nu}(k)=0~,
\eeq
and
\beq
k^\mu{}^\ast\!\epsilon_{\mu\nu}(k)=k^\mu{}^\ast\!\epsilon^A_{\mu\nu}(k)=k^\mu{}^\ast\!\epsilon^B_{\mu\nu}(k)=0~.
\label{eq.bianchi}
\eeq
By contracting $k^\mu$ into Eqs.~\eqref{e.FepsDef} and~\eqref{e.starFepsDef} we find
\begin{align}
k^\mu\epsilon_{\mu\nu}(k)=0=&-\frac{n\cdot k}{n^2}n^\alpha\left[\epsilon^A_{\nu\alpha}(k)+{}^\ast\!\epsilon^B_{\nu\alpha}(k) \right]\;\Rightarrow\;n^\alpha\left[\epsilon^A_{\nu\alpha}(k)+{}^\ast\!\epsilon^B_{\nu\alpha}(k) \right]=0\label{e.ABstarEpsIdent}\\
k^\mu{}^\ast\!\epsilon_{\mu\nu}(k)=0=&-\frac{n\cdot k}{n^2}n^\alpha\left[\epsilon^B_{\nu\alpha}(k)-{}^\ast\!\epsilon^A_{\nu\alpha}(k) \right]\;\Rightarrow\;n^\alpha\left[\epsilon^B_{\nu\alpha}(k)-{}^\ast\!\epsilon^A_{\nu\alpha}(k) \right]=0~.\label{e.BAstarEpsIdent}
\end{align}
In short, the $A$ and $B$ polarization tensors satisfy quite constraining relationships. This should not come as too much of a surprise since all these tensors are describing the same two physical degrees of freedom of the photon.

These relationships allow us to significantly, and remarkably, simplify~\eqref{e.FepsDef}. We find
\begin{align}
\epsilon_{\mu\nu}=&\frac{n^\alpha}{n^2}\left[n_\mu\left(\epsilon^A_{\alpha\nu}+{}^\ast\!\epsilon^B_{\alpha\nu}\right)-n_\nu\left(\epsilon^A_{\alpha\mu}+{}^\ast\!\epsilon^B_{\alpha\mu}\right)\right]-{}^\ast\!\epsilon^{B}_{\mu\nu}\nonumber\\
=&\frac{n^\alpha}{n^2}\left[n_\mu\left(\epsilon^A_{\alpha\nu}-\epsilon^A_{\alpha\nu}\right)-n_\nu\left(\epsilon^A_{\alpha\mu}-\epsilon^A_{\alpha\mu} \right)\right]-{}^\ast\!\epsilon^{B}_{\mu\nu}\nonumber\\
=&-{}^\ast\!\epsilon^{B}_{\mu\nu}~.
\end{align}
Similarly, we find that~\eqref{e.starFepsDef} becomes
\beq
 {}^\ast\!\epsilon_{\mu\nu}={}^\ast\!\epsilon^{A}_{\mu\nu}~.
\eeq
Or, to summarize we have 
\beq
\epsilon_{\mu\nu}=\epsilon^A_{\mu\nu}=-{}^\ast\!\epsilon^{B}_{\mu\nu}~, \ \ \ \ {}^\ast\!\epsilon_{\mu\nu}={}^\ast\!\epsilon^A_{\mu\nu}=\epsilon^{B}_{\mu\nu}~.\label{e.PolRel}
\eeq
These results make clear that each of these polarization tensors contains all the information about the photon degrees of freedom. 

The relations in~\eqref{e.PolRel} also enable us to rewrite the Zwanziger axial anomaly results
\begin{align}
-\frac{e^2q^2}{4\pi^2}\overline{\epsilon}^A_{\mu\nu}(k){}^\ast \overline{\epsilon}^{A\mu\nu}(p)=&-\frac{e^2q^2}{4\pi^2}\overline{\epsilon}_{\mu\nu}(k){}^\ast \overline{\epsilon}^{\mu\nu}(p)\to-\frac{e^2q^2}{8\pi^2}F_{\mu\nu}{}^\ast\!F^{\mu\nu}~,\\
-\frac{b^2g^2}{4\pi^2}\overline{\epsilon}^B_{\mu\nu}(k){}^\ast \overline{\epsilon}^{B\mu\nu}(p)=&\frac{e^2q^2}{4\pi^2}\overline{\epsilon}_{\mu\nu}(k){}^\ast \overline{\epsilon}^{\mu\nu}(p)\to\frac{b^2g^2}{8\pi^2}F_{\mu\nu}{}^\ast\!F^{\mu\nu} ~,\\
-\frac{ebqg}{2\pi^2}\overline{\epsilon}^A_{\mu\nu}(k){}^\ast \overline{\epsilon}^{B\mu\nu}(p)=&\frac{ebqg}{2\pi^2}\overline{\epsilon}_{\mu\nu}(k) \overline{\epsilon}^{\mu\nu}(p)\to\frac{ebqg}{4\pi^2}F_{\mu\nu}F^{\mu\nu} ~.
\end{align}
In other words, the total axial anomaly can be written completely in terms of the physical field strengths. When this is done we find 
\begin{align}
 - \frac{1}{8\pi^2}(e^2q^2 - b^2 g^2)F_{\mu\nu}{}^\ast\!F^{\mu\nu}
   +\frac{ebqg}{4\pi^2} F_{\mu\nu}F^{\mu\nu} ~.
\end{align}
This shows that calculations in terms of the Zwanziger field strengths $F^A_{\mu\nu}$ and $F^B_{\mu\nu}$ produce the same results as obtained using the physical field strength $F_{\mu\nu}$.

%%%%%%%%%%%%%%%%%%%%%%%%%%%%%%%%%%%%%%%%%%%%%%%%%%%%%%%%%%%%
\section{Conclusion\label{s.con}}
This article provides a better understanding of how $U(1)$ gauge theories can be described in a gauge invariant way. The field strength, rather than the potential, is used to describe the photon, but couples to a surface bounded by the current worldline rather than the current itself. Our previous work indicated that the ambiguity of gauge transformations that appears in the potential formalism is replaced by a geometrical ambiguity, that of which Stokes surface to use, when using these field strength methods.

In this work we make this connection firm. We have shown that surface invariance is tightly linked the Ward identity. That is, when a photon coupling in an amplitude  satisfies the Ward identity then the corresponding  field strength coupling in the amplitude is surface independent. This general point is explored in an interesting and nontrivial way by considering the gauge anomalies of $U(1)$ gauge theories.

We showed how the gauge anomaly arises in this gauge invariant field strength formalism as a dependence on arbitrary Stokes surfaces. That is, in this gauge invariant formalism these is no way to express the three photon amplitude in a surface independent way. This contrasts with potential methods in which there is no way to express the amplitude in a gauge invariant way.

 We have also explicitly calculated the contribution of electric and magnetic particles to the axial anomaly using the field strength formalism. We find consistency with calculations based on duality and also those using the Zwanziger formalism. It would be interesting to extend the field strength formalism to loop calculations with internal photon lines and also to a more general study of $\theta$ terms in Abelian gauge theories with multiple gauge fields.
 
There is also a suggestive similarity between these gauge invariant methods and the gauge invariant spinor-helicity formalism. The field strength methods introduce vectors that define the surfaces the field strengths couples to. While the final results cannot depend on the choice of surface, and hence vector, some choice must be made for the calculation to be made at all. Spinor-helicity methods also introduce reference spinors/vectors in order to calculate. Again, the final results do not depend on the particular choice made, but the reference spinors are essential to making the calculations. It would be interesting to see if this calculational similarity hints at a deeper connection between these two, seemingly quite different, formalisms.

\section*{Acknowledgments}
We thank Csaba Csáki, Rotem Ovadia, Maximilian Ruhdorfer, and Ofri Telem for enlightening discussions. J.T.  is supported in part by the DOE under grant DE-SC-0009999. C.B.V. is supported in part by the National Science Foundation under Grant No. PHY-2210067~.

\appendix

\end{document}